\documentclass[twocolumn,showpacs,superscriptaddress,prb,amsmath,amssymb,floatfix]{revtex4}
\usepackage{times}
\usepackage{amsmath,bm,amsfonts}
\usepackage{dcolumn}
\usepackage{graphicx}
\usepackage{latexsym}

\begin{document}

\title{Doped Valence-bond Solid and Superconductivity on the
Shastry-Sutherland Lattice}

\author{Bohm-Jung \surname{Yang}}

\affiliation{Department of Physics and Astronomy, Center for Strongly
Correlated Materials Research, and Center for Theoretical Physics,
Seoul National University, Seoul 151-747, Korea}

\author{Yong Baek \surname{Kim}}

\affiliation{Department of Physics, University of Toronto,
Toronto, Ontario M5S 1A7, Canada}
\affiliation{School of Physics,
Korea Institute for Advanced Study, Seoul 130-722, Korea}

\author{Jaejun \surname{Yu}}

\affiliation{Department of Physics and Astronomy, Center for
Strongly Correlated Materials Research, and Center for Theoretical
Physics, Seoul National University, Seoul 151-747, Korea}

\author{Kwon \surname{Park}}
\email{kpark@kias.re.kr}

\affiliation{School of Physics, Korea Institute for Advanced
Study, Seoul 130-722, Korea}

\date{\today }

\begin{abstract}
Motivated by recent experiments on SrCu$_2$(BO$_3$)$_2$, we
investigate the ground states of the doped Mott insulator on the
Shastry-Sutherland lattice. To provide a unified theoretical
framework for both the valence-bond solid state found in undoped
SrCu$_2$(BO$_3$)$_2$ and the doped counterpart being pursued in on-going
experiments, we analyze the $t$-$J$-$V$ model via the bond
operator formulation. It is found that novel superconducting states
emerge upon doping with their properties crucially depending on
the underlying valence bond order. Implications to future
experiments are discussed.
\end{abstract}


\maketitle

\section{\label{sec:intro} Introduction}

The fate of doped Mott insulators is one of the most challenging
issues in correlated electron physics, especially in relation to
the long-standing problem of high-temperature superconductivity in
cuprates \cite{sachdevrev,kivelson,palee}. Recent discoveries of
various Mott insulators on geometrically frustrated lattices
\cite{Kageyama,hiroi,yslee,takagi} and organic materials
\cite{kanoda} may offer an important clue to this issue when such
materials are doped. While it is possible that the ground state
does not break any symmetry, resulting in a spin liquid phase, the
ground states of Mott insulators often have broken spin-rotation
and lattice-translation symmetries, leading to antiferromagnetic
and valence bond solid order, respectively \cite{sachdevrev}.
Since both antiferromagnetic and valence bond solid phases are
generic possibilities for Mott insulators, a zero-temperature
quantum phase transition may occur between these two phases when
an appropriate ``control parameter'' is changed \cite{sachdevrev}.
In this context, understanding the effect of doping on
valence-bond solid phases is as equally important as that on
antiferromagnetic phases and would be quite useful for the full
classification of all possible phases of doped Mott insulators.

There are, however, not many clear examples of two-dimensional
valence-bond solid insulator in contrast to the antiferromagnetic
insulator found in high $T_c$ cuprate compounds. The discovery of
SrCu$_2$(BO$_3$)$_2$ is particularly important in this regard
\cite{Kageyama,kodama,Gaulin}. This material can be characterized
by an antiferromagnetic spin-1/2 Heisenberg model on the
Shastry-Sutherland lattice \cite{ss1}. Starting from the usual
square lattice, the Shastry-Sutherland lattice can be obtained by
putting additional diagonal bonds with two possible orientations
in alternating plaquettes (see Fig.1). Let $J$ and $J'$ be the
exchange couplings along the diagonal and the square lattice
links, respectively. It is known that the valence bond solid state
or the product state of valence bond singlets (illustrated as
filled ellipses in Fig.1) along the diagonal bonds is the exact
ground state of the Heisenberg model for $J'/J < 0.7$
\cite{ss1,mila, ueda1, Weihong, Muller1,Koga,Sigrist,Chung}. In
SrCu$_2$(BO$_3$)$_2$,  $J'/J$ is estimated to be 0.64
\cite{Kageyama}.

\begin{figure}[t]
\centering
\includegraphics[width=7.5cm]{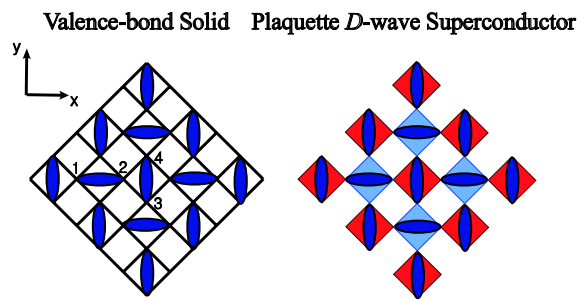}
\caption{(Color online) Schematic diagram for the ground states on
the Shastry-Sutherland lattice. The valence-bond solid state is
the ground state at half filling as depicted in the left figure
where ellipses represent spin-singlet pairs. When doped with
holes, the system exhibits superconductivity in addition to the
coexisting valence-bond solid order. Furthermore, when the
nearest-neighbor repulsive interaction, $V$, is larger than a
critical value, $V_c$, the plaquette $D$-wave superconductivity
appears in a range of doping concentration, $x$, with a peculiar
spatial pattern as shown in the right figure.  In this situation,
the pairing amplitudes residing in the four links encircling the
horizontal dimers have the opposite sign to those for the vertical
dimers. Different colors are used to emphasize the plaquette
pattern of the pairing amplitude.} \label{fig:1}
\end{figure}

In this paper, we investigate possible phases of doped
valence-bond solid and superconductivity on the Shastry-Sutherland
lattice. In order to study the interplay between the
valence-bond solid order and emergent superconductivity at finite
doping, we use the bond operator formulation
\cite{kpark1,sachdev,kpark2} of the constrained Hilbert space of
correlated electron systems, extended to general doping and
applied to the $t$-$J$-$V$ model. Here $t$ and $V$ represent the
hopping strength and nearest-neighbor repulsion between electrons,
respectively. In contrast to previous studies of the
Shastry-Sutherland model \cite{ss2,ybkim,Liu}, the emergent
superconducting state is directly related to the underlying
valence-bond solid order at the half-filling. For example, the
valence-bond solid order is so robust that superconductivity
always coexists with it as shown in the resulting phase diagram of
Fig.~2.

When the nearest-neighbor repulsion, $V$, is smaller than a
critical value $V_c$, the doped holes can remian paired mostly
{\it within} each valence-bond singlet or dimer on the diagonal
bonds. This leads to $S$-wave superconductivity. On the other
hand, when $V$ is larger than $V_c$, the Coulomb repulsion
prevents doped holes from occupying the same dimer. At larger
doping concentrations, the doped holes tend to develop pairing
correlation between nearby dimers of the valence-bond singlets.
Consequently the superconducting order paramter acquires an
interesting phase relation on the square lattice link, depending
on whether the square lattice links encircle the horizontal or vertical
dimers (See Fig.~1).
More specifically, all the links within the same plaquette have
the same sign, but those in the nearest-neighbor plaquettes have
the opposite sign. Since this looks like $D$-wave symmetry at long
distances, we call it the {\it plaquette} $D$-wave
superconductivity. Notice, however, that this is different from
the ordinary $d$-wave superconductor on the square lattice where
the sign alternates for different links within the same plaquette.
This peculiar structure comes from the underlying valence bond
solid order. The simultaneous presence of the valence bond order
and superconductivity can, in principle, be checked by X-ray or
neutron scattering experiments.

The rest of the paper is organized as follows. In section
\ref{sec:method}, we review the bond operator formulation for
doped magnets. In section \ref{sec:hamiltonian}, the $t$-$J$-$V$
Hamiltonian is written in the bond operator representation and the
mean field theory is explained. In section \ref{sec:results}, the
results of the mean field theory and the phase diagram are
discussed. We conclude in section \ref{sec:conclusion}. Some
details of the computations are relegated to the appendix.

\begin{figure}[t]
\centering
\includegraphics[width=7.5cm]{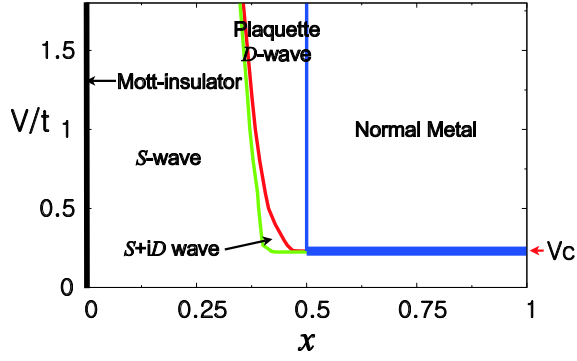}
\caption{(Color online) Phase diagram of the $t$-$J$-$V$ model on
the Shastry-Sutherland lattice as a function of hole
concentration, $x$, and the nearest-neighbor repulsive
interaction, $V$. The thick line separating the normal metal and
the $S$-wave superconducting phase is a first-order phase
boundary, while other boundaries are all second-order. The undoped
Mott-insulating state is represented by the thick solid line at
$x=0$.} \label{fig:2}
\end{figure}

\section{\label{sec:method} Bond Operator Formalism}

We start with the bond operator theory by setting up an exact
mapping between bond operators and usual electron creation
operators \cite{sachdev,kpark1,kpark2}. Let $c_{1a}^{\dag}$ and
$c_{2a}^{\dag}$ ($a=\uparrow , \downarrow $) be the electron
creation operators on the left and right site, respectively, of
the lattice link that a horizontal dimer occupies (see
Fig.~\ref{fig:1}). In the limit of large on-site repulsive
interaction, $U$, any state with two electrons at the same site is
excluded from the low-energy Hilbert space which in turn can be
represented by nine ``bond-particle'' creation operators defined
as follows:

(i) Singlet boson for spin-Peierls order,
\begin{equation}
s^{\dagger} |v\rangle  =\frac{1}{\sqrt{2}}
\varepsilon_{ab}c_{1a}^{\dag}c_{2b}^{\dag} |0\rangle ,
\end{equation}
where $|0\rangle$ is the electron vacuum and $|v\rangle$ is the
imaginary vacuum void of any bond particle. $\varepsilon_{ab}$ is
the second-rank antisymmetric tensor with $\varepsilon_{\uparrow
\downarrow}=1$. From now on we adopt the summation
convention for repeated spin indices (such as $a$ and $b$
above) throughout this paper unless mentioned otherwise.

(ii) Triplet magnon,
\begin{equation}
t^{\dagger}_{\alpha} |v\rangle  =\frac{1}{\sqrt{2}}
\sigma^{\alpha}_{bc}\varepsilon_{ca} c_{1a}^{\dag}c_{2b}^{\dag}
|0\rangle,
\end{equation}
where $\sigma^{\alpha}_{ab}$ $(\alpha=x,y,z)$ are the
usual Pauli matrices.

(iii) Single-hole fermion,
\begin{align}
h^{\dagger}_{1a} |v\rangle &= c_{1a}^{\dagger} |0\rangle,
\nonumber\\
h^{\dagger}_{2a} |v\rangle &= c_{2a}^{\dagger} |0\rangle.
\end{align}

(iv) Double-hole boson for the empty state,
\begin{equation}\label{eq:d}
d^{\dagger} |v\rangle = |0\rangle.
\end{equation}

As indicated by the names, the operators $s$, $d$, and
$t_{\alpha}$ all obey the canonical boson commutation relations
while $h_{1a}$ and $h_{2a}$ satisfy the canonical fermion
anticommutation relations. To eliminate unphysical states from the
enlarged Hilbert space, the following constraint needs to be
imposed on the bond-particle Hilbert space:
\begin{equation} \label{eq:constraint}
s^{\dagger}s +t_{\alpha}^{\dagger}t_{\alpha} +
h_{1a}^{\dagger}h_{1a} +h_{2a}^{\dagger}h_{2a} +d^{\dag}d = 1.
\end{equation}
Constrained by this equation,
the exact expressions for
the spin and electron creation operators can be written in terms of
the bond operators. For example,
\begin{align}\label{eq:bond-ops}
S_{1\alpha}&=\frac{1}{2}(s^{\dag}t_{\alpha} +t_{\alpha}^{\dag}s
-i\varepsilon_{\alpha\beta\gamma}t_{\beta}^{\dag}t_{\gamma})
+\frac{1}{2}\sigma^{\alpha}_{ab}h^{\dag}_{1a}h_{1b},
\nonumber\\
S_{2\alpha}&=-\frac{1}{2}(s^{\dag}t_{\alpha} +t_{\alpha}^{\dag}s
+i\varepsilon_{\alpha\beta\gamma}t_{\beta}^{\dag}t_{\gamma})
+\frac{1}{2}\sigma^{\alpha}_{ab}h^{\dag}_{2a}h_{2b},
\nonumber\\
c^{\dag}_{1a}&=h^{\dag}_{1a}d
+\frac{1}{\sqrt{2}}\varepsilon_{ab}s^{\dag}h_{2b}
-\frac{1}{\sqrt{2}}\varepsilon_{ac}\sigma^{\alpha}_{cb}t^{\dag}_{\alpha}h_{2b},
\nonumber\\
c^{\dag}_{2a}&=h^{\dag}_{2a}d
+\frac{1}{\sqrt{2}}\varepsilon_{ab}s^{\dag}h_{1b}
+\frac{1}{\sqrt{2}}\varepsilon_{ac}\sigma^{\alpha}_{cb}t^{\dag}_{\alpha}h_{1b},
\end{align}
where $\varepsilon_{\alpha\beta\gamma}$ is the third rank
antisymmetric tensor with $\varepsilon_{xyz}=1$.

Now, in order to cover the full Shastry-Sutherland lattice, we
need nine additional bond operators representing the electronic
Hilbert space associated with the vertical dimers corresponding to the
sites 3 and 4. For example, $\tau_{\alpha}$ indicates the triplet
magnon in the vertical dimers while $h_{3a}$ and $h_{4a}$ represent
the corresponding single-hole fermions.
While non-uniform condensations of the spin-Peierls singlet and/or
double-hole bosons are in general possible, we focus only on the
uniform phases in this study. Thus the condensation densities of
the singlet and double-hole bosons will be represented by
$\bar{s}$ and $\bar{d}$ for both horizontal and vertical dimers.

\section{\label{sec:hamiltonian} Choice of Hamiltonian and Bond Operator Mean Field Theory}

We consider the following $t$-$J$-$V$ Hamiltonian:
\begin{align}
H = \hat{P}_\textrm{G} \Big[ &-\sum_{\langle i,j
\rangle}t_{ij}(c^{\dag}_{ia}c_{ja}
+\textrm{H. c.}) -\mu \sum_i c^{\dag}_{ia} c_{ia} \nonumber \\
&+\sum_{\langle i,j \rangle} J_{ij} {\bf S}_i \cdot {\bf S}_j
+\sum_{\langle i,j \rangle} V_{ij} n_i n_j \Big]
\hat{P}_\textrm{G} \;,
\end{align}
where $\hat{P}_\textrm{G}$ is the Gutzwiller projection operator
imposing the no-double-occupancy constraint. $t_{ij}$ and $J_{ij}$
are the electron hopping matrix element and antiferromagnetic
exchange interaction, respectively. In this paper, we set
$t_{ij}=t$ and $J_{ij}=J$ within a dimer, $t_{ij}=t'$ and
$J_{ij}=J'$ between neighboring dimers. We take $J'/J=0.64$ from
the experiments on SrCu$_2$(BO$_3$)$_2$ \cite{Kageyama,ss2}. From
the large-$U$ expansion of the Hubbard model, we can then set
$t'/t= \sqrt{J'/J}=0.8$. However, the parameter, $J'/t'$, which is
important for hole dynamics, is not available experimentally. Thus
we follow the convention used in previous works \cite{ss2} and
take $J'/t'=0.3$. Also, $\mu$ is the chemical potential and $n_i$
is the electron density operator. Finally, for convenience, the
nearest-neighbor repulsive interaction ,$V_{ij}$, is set to $V$
for both the two sites within dimer and those between the
nearest-neighbor dimers.

We now write the
$t$-$J$-$V$ Hamiltonian solely in terms of the bond particle
operators \cite{kpark1}. As usual, the constraint on the bond
particle operators is imposed by the Lagrange multiplier
method. Residual interactions between bond particles are analyzed
via quadratic decoupling of quartic terms in a similar manner to
the usual Hartree-Fock-BCS treatment. For the mean-field
description, we consider the following order parameters:
\begin{align}\label{eq:orderparameter}
P_{x} &\equiv \langle t^{\dag}_{{\bf i}\alpha} \tau_{{\bf
i}+\hat{x},\alpha} \rangle, \; Q_{x} \equiv \langle t_{{\bf
i}\alpha} \tau_{{\bf i}+\hat{x},\alpha} \rangle,
\nonumber\\
\Pi_{x} &= \langle h^{\dag}_{1{\bf i}a} h_{3,{\bf i}-\hat{x},a}
\rangle = \langle h^{\dag}_{1{\bf i}a} h_{4,{\bf i}-\hat{x},a}
\rangle
\nonumber\\
&= \langle h^{\dag}_{2{\bf i}a} h_{3,{\bf i}+\hat{x},a} \rangle =
\langle h^{\dag}_{2{\bf i}a} h_{4,{\bf i}+\hat{x},a} \rangle,
\nonumber\\
\Delta_{x} &= \langle h_{1{\bf i}\downarrow} h_{3,{\bf
i}-\hat{x},\uparrow} \rangle = \langle h_{1{\bf i}\downarrow}
h_{4,{\bf i}-\hat{x},\uparrow} \rangle
\nonumber\\
&= \langle h_{2{\bf i}\downarrow} h_{3,{\bf i}+\hat{x},\uparrow}
\rangle = \langle h_{2{\bf i}\downarrow} h_{4,{\bf
i}+\hat{x},\uparrow} \rangle,
\end{align}
where ${\bf i}$ is the dimer index of the horizontal dimers and
${\bf i}\pm\hat{x}$ indicates the locations of the neighboring
vertical dimers. The order parameters for the $y$-direction can be
defined similarly.

We consider the possibility that the valence-bond-solid order
persists even at non-zero doping. We further consider the
condensation of $d$ bosons, but neglect the possibility of triplet
condensation because we are mostly concerned about paramagnetic
phases in this work. All of the order parameters ($P$, $Q$, $\Pi$,
and $\Delta$ for both $x$ and $y$ directions) defined above as
well as $\bar{s}$, $\bar{d}$, the chemical potential, $\mu$, and
the Lagrange multiplier, $\xi$, are determined by solving a
coupled set of four saddle-point equations and eight
self-consistency equations. Details of the computational procedure
are presented in the Appendix.

\section{\label{sec:results} Results of the Mean Field Theory and the Phase Diagram}

At zero doping, the system is in a robust valence-bond solid
phase. In our bond operator theory, the robustness of the ground
state can be seen by the complete localization of triplet magnons:
the $t_\alpha$ and $\tau_\alpha$ dispersions are completely flat
and high in energy. It can easily be shown that, at the quadratic
order, the triplet magnons on the Shastry-Sutherland lattice are
completely decoupled from the singlet contributions (Notice that
the coupling between the singlets and triplets at the quadratic order
were the main driving force for the triplet dispersion in the square-lattice
case \cite{kpark1}). This leads to the triplet Hamiltonian with
no dispersive quadratic part. Thus, the only way to generate
the triplet dispersion is through the saddle-point order
parameters, $P_{x}$, $P_{y}$,$Q_{x}$ and $Q_{y}$.
It turns out, however, that all of the above order parameters are
actually zero for any $J'/J$ at half filling in our mean-field
theory so that the triplets are completely localized.

\begin{figure}[t]
\centering
\includegraphics[width=8cm]{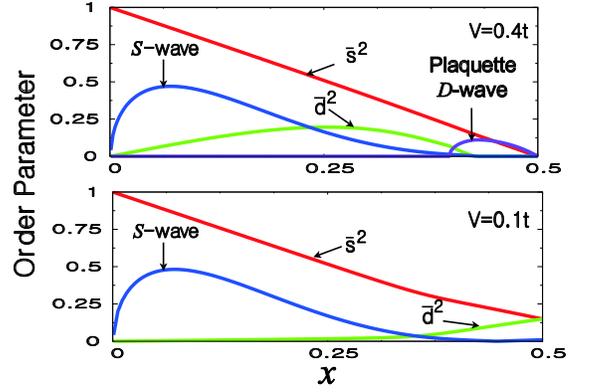}
\caption{(Color online) Superconducting order parameter, $\chi$,
as well as condensation densities of the spin-Peierls singlets,
$\bar{s}$, and double-hole bosons, $\bar{d}$, as a function of
hole concentration, $x$. When $V=0.4t>V_{c}$ (top panel), $S$-wave
$\chi$ and $\bar{d}$ simultaneously vanish at $x=0.431$. Before
the full plaquette $D$-wave superconductivity sets in at larger
$x$, there is a region, $0.389<x<0.431$, where the $S$- and
$D$-wave superconductivity coexists in the form of $S$+i$D$-wave.
Finally at $x=0.5$, the plaquette $D$-wave superconducting order
parameter becomes zero and so does $\bar{s}$. No valence-bond
solid correlation exists for $x \geq 0.5$. When $V=0.1t<V_{c}$
(bottom panel), $\bar{d}$ increases monotonically as a function of
$x$ while $\bar{s}$ remains finite all the way to $x=1$. For
clarity the $S$- and plaquette $D$-wave superconducting order
parameters as well as $\bar{d}^2$ are magnified twenty times in
the top panel while, in the bottom panel, only the $S$-wave
superconducting order parameter is magnified twenty times.}
\label{fig:3}
\end{figure}
Now let us consider the case of non-zero doping. In our
Hartree-Fock-BCS saddle-point approximation, superconductivity is
found to appear at any non-zero doping when one neglects long-rage
charge inhomogeneities such as Wigner crystal order at very small $x$.
In the bond operator representation, the electron superconducting
order parameter can be computed from
\begin{align}
\chi_{x} &= \varepsilon_{ab} \langle c^{\dag}_{1{\bf i}a}
c^{\dag}_{3,{\bf i}-\hat{x},b} \rangle = \varepsilon_{ab} \langle
c^{\dag}_{1{\bf i}a} c^{\dag}_{4,{\bf i}-\hat{x},b} \rangle
\nonumber\\
&= \varepsilon_{ab} \langle c^{\dag}_{2{\bf i}a} c^{\dag}_{3,{\bf
i}+\hat{x},b} \rangle = \varepsilon_{ab} \langle c^{\dag}_{2{\bf
i}a} c^{\dag}_{4,{\bf i}+\hat{x},b} \rangle
\nonumber\\
&= (2\bar{d}^2-\bar{s}^2-Q_{x})\Delta_{x}
-\sqrt{2}\bar{s}\bar{d}\Pi_{x} \;. \label{chi}
\end{align}
The superconducting order parameter for the $y$-direction is
defined similarly. The doping dependence of $\chi$ is plotted in
Fig. \ref{fig:3} along with the singlet boson condensate density,
$\bar{s}$, and the double-hole boson condensate density,
$\bar{d}$.

\begin{figure}[t]
\centering
\includegraphics[width=7cm]{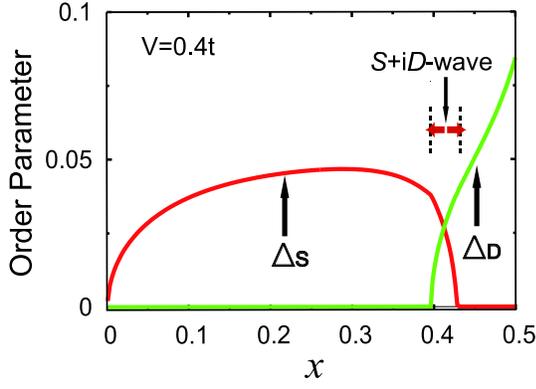}
\caption{$h$-fermion pairing amplitude, $\Delta$, as a function of
hole concentration, $x$. In the plot, $\Delta_{S}$ and
$\Delta_{D}$ indicate the $S$- and $D$-wave components of $\Delta$,
respectively. Note that there is a region, $0.389<x<0.431$, where
the $S$- and $D$-wave superconductivity coexist in the form of the
$S$+i$D$-wave.} \label{fig:4}
\end{figure}

While $\chi$ can be a complex number in general, it turns out to
be real for most of the phase space, barring an
arbitrary overall phase factor (See Fig.~\ref{fig:2}). In
Fig.~\ref{fig:2}, however, there is a narrow region where the $S$-
and $D$-wave superconductivity coexists in the form of
an $S$+i$D$-state similar to the results of previous studies
on the square lattice \cite{vojta}. In this case $\chi_x$ is a complex number and
$\chi_x = \chi^*_y $. As shown in Eq.~(\ref{chi}), the $h$-fermion
pairing amplitude, $\Delta$, is the key element in determining the
sign of $\chi$ and thereby the pairing symmetry of the electronic
superconducting order parameter. Generally the momentum dependence
of $\Delta$ can be written as follows:
\begin{align}
\Delta_{S+\textrm{i}D} &= \Delta_x \cos{k_{x}}+\Delta_y
\cos{k_{y}}
\nonumber \\
&=\Delta e^{i \varphi}
\cos{k_{x}}+\Delta e^{-i \varphi}\cos{k_{y}}
\nonumber\\
&\equiv\Delta_{S}(\cos{k_{x}}+\cos{k_{y}})
+i\Delta_{D}(\cos{k_{x}}-\cos{k_{y}}).
\end{align}
If both $\Delta_{S}$ and $\Delta_{D}$ remain finite, the system
possesses the $S$+i$D$-wave pairing symmetry. In Fig.~\ref{fig:4}
we plot the detailed doping dependence of $\Delta_{S}$ and
$\Delta_{D}$ for $V=0.4t$.

Notice that the critical exponent, $\beta$, defined in $\chi
\sim|x-x_{c}|^{\beta}$ shows an interesting behavior when
$V>V_{c}$. Superconducting order parameters for both $S$- and
$D$-wave show the usual mean-field behavior, i.e., $\beta=1/2$,
when they first emerge. On the other hand, when they disappear,
both exhibit an unusual exponent of $\beta=1$. This deviation from
the conventional mean-field behavior has different origins in each
case. For the $S$-wave case, the phase boundary corresponds to the
critical point between $S$+i$D$-wave and $D$-wave
superconductivity, where the presence of nodal fermions leads to a
non-analytic cubic term in the expansion of the ground state
energy \cite{sachdev2,Laughlin}. On the other hand, for the
$D$-wave, the linearly vanishing $\bar{s}^2$ gives rise to
$\beta=1$ (Note that $\Delta_D$ remains finite at the critical
point $x=0.5$. See Fig.~\ref{fig:4} ).

It is also interesting to note that our $h$-fermion (defined on a
dimer) is an extended object which carries both the charge and
spin quantum numbers. As a consequence, when the $h$-fermion
pairing amplitude is finite, the distance between two holes in a
$h$-$h$ pair is larger than the average electron distance, as
reported in exact diagonalization studies \cite{Leung}. This
behavior is different from what one would expect in the
slave-boson-type theory \cite{ss2,ybkim}.

In Fig.~\ref{fig:2} the zero-temperature phase diagram is plotted
as a function of hole concentration, $x$, and the nearest-neighbor
repulsive interaction, $V$. The overall shape of the phase diagram
is determined by the behavior of the $d$-boson condensate, which
corresponds to the local pairing of holes within dimers. For
$V<V_{c}\simeq 0.23t$, the $d$-boson condensation is the primary
mechanism of pairing and its short-range nature generates $S$-wave
pairing rather than $D$-wave with a nodal structure. On the other
hand, for $V>V_{c}$, there is a $V$-dependent critical hole
concentration, $x_c$, where $\bar{d}$ vanishes. It is this
collapse of $d$-boson condensates that makes pairing more
long-ranged and finally leads to the emergence of $D$-wave
superconductivity.

The detailed $x$-dependence of the superconducting order
parameter, $\chi$, is shown in Fig.~\ref{fig:3}. For $V=0.4t>V_c$,
the $S$-wave superconducting order parameter and the $d$-boson
condensate density go to zero simultaneously at $x \simeq 0.431$.
The collapse of the $d$-boson condensate at finite doping, in
turn, imposes a precise upper bound on the hole concentration, up
to which the valence-bond order can exist. To see this, note that,
in the bond-operator representation, the chemical potential,
$\mu$, is fixed to satisfy $ \langle h^{\dag}_{1{\bf i}a} h_{1{\bf
i}a} +h^{\dag}_{2{\bf i}a} h_{2{\bf i}a} \rangle +2\bar{d}^2 =
2x$. A similar equation exists for $h_3$ and $h_4$ fermions. When
$\bar{d}^2=0$, hole doping can be achieved only through
$h$-fermions. In this situation, the hole concentration cannot
exceed $x=1/2$ where every dimer has exactly one electron. Note
that, for $V=0.1t<V_c$, $\bar{d}^2$ increases monotonically as $x$
increases and finally reaches unity at $x=1$.

The plaquette $D$-wave superconducting order parameter found here
has a peculiar spatial pattern as shown in Fig.~\ref{fig:1}. As
explained in the introduction, this pattern is completely
different from the corresponding pattern of the conventional
$d$-wave superconducting state that previous slave-boson theories
would predict \cite{ss2,ybkim}. We argue that the order parameter
pattern of our $D$-wave state is rather natural in the limit of
strong dimerization. In this limit, two electrons within the same
dimer lose their independent local coordinates and share the same
single dimer index. Then, the links of the plaquette enclosing the
horizontal (vertical) dimers would connect only the links of the
nearest-neighbor plaquettes that encircle the vertical
(horizontal) dimers. Thus it is natural to expect the
plaquette-dependent order parameter.

\begin{figure}[t]
\centering
\includegraphics[width=7cm]{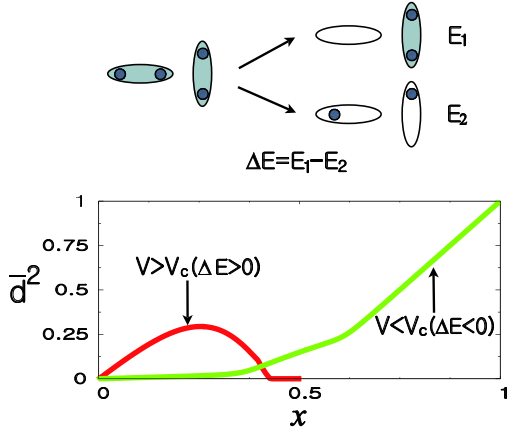}
\caption{(Color online) Schematic diagram showing the origin of
different behaviors of the $d$ boson condensate density below and
above $V_c \simeq 0.23t$. The bottom figure plots $\bar{d}^2$ as a
function of hole concentration, $x$, for $ V=0.4t$ ($\Delta E>0$)
and $V=0.1t$ ($\Delta E<0$). Note that for clarity $\bar{d}^2$ is
magnified thirty times in the case of $V=0.4t$.} \label{fig:5}
\end{figure}

Completely different behaviors of the $d$-boson condensate density
below and above $V_c$ can physically be understood as follows.
When two holes are doped into a unit cell, there are two possible
configurations as shown in Fig.~\ref{fig:5}. First, two electrons
are removed from the same dimer while the other remains intact. In
this case, only one singlet pair is removed, which costs the
energy of $E_1 = E_s \sim J$ with an additional Coulomb energy
gain due to the disappearance of nearest-neighboring electrons:
$E_d \sim V$. Thus the total energy cost would be $E_1 = E_s -E_d
\sim J-V$. Second, when two electrons are removed from two dimers,
i.e., one hole remains at each dimer, there are twice the Coulomb
energy gain and twice the exchange energy cost for breaking two
singlet pairs. The total energy cost for the second case is $E_2
\sim 2 E_1$. In conclusion, the energy-cost difference between two
configurations is $\Delta E = E_1 - E_2 \sim V-J$. If $\Delta E <
0$, the first configuration is preferred and the $d$ bosons can
condense. On the other hand, if $\Delta E > 0$, the second
configuration is favored and the $d$ bosons may not condense
depending on $x$, which is precisely the case as shown in the
bottom figure of Fig.~\ref{fig:5}.

\section{\label{sec:conclusion} Conclusion}

By analyzing the $t$-$J$-$V$ model based on the bond operator
formalism, we have investigated the phase diagram of the doped
Mott insulator on the Shastry-Sutherland lattice. The interplay
between strong dimerization and nearest-neighbor repulsive
interactions leads to different behaviors of the doped holes
determining the overall phase diagram. If the nearest-neighbor
repulsive interaction, $V$, is smaller than a critical value,
$V_c$, the hole-pairing within dimers is preferred, resulting in
$S$-wave superconductivity at any non-zero $x$. On the other hand,
if $V$ is larger than $V_c$, the density of paired-holes within
the same dimers vanishes at finite $x$ and the plaquette $D$-wave
superconductivity with a peculiar spatial pattern emerges.

Considering clear experimental evidence for the valence-bond solid
state in undoped SrCu$_2$(BO$_3$)$_2$, we believe that the above
conclusions would be valid for a range of realistic situations. It
will be interesting to observe the possible transition between the
$S$- and $D$-wave superconducting states.
Finally we note that, while doping mobile carriers to
SrCu$_2$(BO$_3$)$_2$ has not been achieved yet, there has been a
recent progress in doping quenched non-magnetic Mg impurities to
this material \cite{Mg_doping}.

This work was supported by the NSERC, CRC, CIAR,
KRF-2005-070-C00044 (YBK) and by the KOSEF through CSCMR SRC (BJY, JY).

\appendix
\section{Computational details}
\label{appendix:details}

Here we present details of the Hartree-Fock-BCS mean field theory
in the bond operator formalism. We begin by explicitly writing the
$t$-$J$-$V$ Hamiltonian as follows:
\begin{equation}
H=\hat{P}_\textrm{G} \left( H_{t} +H_{J} +H_{\mu} +H_{V} \right)
\hat{P}_\textrm{G} ,
\end{equation}
where
\begin{align}
H_{t}&=-t \sum_i (
c^{\dag}_{1ia}c_{2ia}+c^{\dag}_{3ia}c_{4ia}+\textrm{H. c.} )
\nonumber\\
&-t' \sum_i \Big[
c^{\dag}_{1ia}(c_{3,i-\hat{x},a}+c_{4,i-\hat{x},a})
\nonumber\\
&\qquad+c^{\dag}_{2ia}(c_{3,i+\hat{x},a}+c_{4,i+\hat{x},a})
+\textrm{H. c.} \Big]
\nonumber\\
&-t'\sum_i \Big[
c^{\dag}_{4ia}(c_{1,i+\hat{y},a}+c_{2,i+\hat{y},a})
\nonumber\\
&\qquad+c^{\dag}_{3ia}(c_{1,i-\hat{y},a}+c_{2,i-\hat{y},a})
+\textrm{H. c.} \Big] ,
\end{align}
\begin{align}
&H_{J}=J\sum_i ( {\bf S}_{1i}\cdot {\bf S}_{2i}+ {\bf S}_{3i}\cdot
{\bf S}_{4i} )
\nonumber\\
&+J'\sum_i \Big[ {\bf S}_{1i}\cdot ({\bf S}_{3,i-\hat{x}}+{\bf
S}_{4,i-\hat{x}}) +{\bf S}_{2i}\cdot ({\bf S}_{3,i+\hat{x}}+{\bf
S}_{4,i+\hat{x}}) \Big]
\nonumber\\
&+J'\sum_i \Big[ {\bf S}_{4i}\cdot ({\bf S}_{1,i+\hat{y}}+{\bf
S}_{2,i+\hat{y}}) +{\bf S}_{3i}\cdot ({\bf S}_{1,i-\hat{y}}+{\bf
S}_{2,i-\hat{y}}) \Big] ,
\end{align}
\begin{align}
H_{\mu}&= -\mu \sum_i ( n_{1i} +n_{2i} -2 +2x) \nonumber \\
&-\mu \sum_i ( n_{3i} +n_{4i} -2 +2x) ,
\end{align}
and
\begin{align}
&H_{V}=V\sum_i ( n_{1i}n_{2i}+ n_{3i}n_{4i} )
\nonumber\\
&+V\sum_i \Big[ n_{1i} (n_{3,i-\hat{x}}+n_{4,i-\hat{x}})+n_{2i}
(n_{3,i+\hat{x}}+n_{4,i+\hat{x}}) \Big]
\nonumber\\
&+V\sum_i \Big[ n_{4i} (n_{1,i+\hat{y}}+n_{2,i+\hat{y}})+n_{3i}
(n_{1,i-\hat{y}}+n_{2,i-\hat{y}}) \Big] .
\end{align}
In the next section, the $t$-$J$-$V$ Hamiltonian is expressed
solely in terms of bond particles.

\subsection{Hamiltonian in the bond-operator representation}

Taking usual steps in any saddle-point theory, we first replace
the full Gutzwiller projection by adding the Lagrange multiplier
term. In the bond operator representation, this Lagrange
multiplier term
is written as follows:
\begin{align}
H_{\xi}&=-\xi \sum_i
(s^{\dag}_{i}s_{i}+t^{\dag}_{i\alpha}t_{i\alpha}
+h^{\dag}_{1ia}h_{1ia}+h^{\dag}_{2ia}h_{2ia} +d^{\dag}_{i}d_{i}-1)
\nonumber\\
&-\xi \sum_i
(\sigma^{\dag}_{i}\sigma_{i}+\tau^{\dag}_{i\alpha}\tau_{i\alpha}
+h^{\dag}_{3ia}h_{3ia}+h^{\dag}_{4ia}h_{4ia}
+\delta^{\dag}_{i}\delta_{i}-1) , \label{H_xi}
\end{align}
where $\xi$ is the Lagrange multiplier. Also, $\sigma_i$ and
$\delta_i$ represent the singlet boson and double-hole boson
operators for the $i$-th vertical dimer, respectively. Other
operators are defined in the main text. Under the constraint
imposed by Eq.~(\ref{H_xi}), the usual chemical potential term can
be written as follows:
\begin{align}
H_{\mu}&= \mu \sum_i (h^{\dag}_{1ia}h_{1ia}+h^{\dag}_{2ia}h_{2ia}
+2d^{\dag}_{i}d_{i}-2x )
\nonumber\\
&+\mu \sum_i (h^{\dag}_{3ia}h_{3ia}+h^{\dag}_{4ia}h_{4ia}
+2\delta^{\dag}_{i}\delta_{i}-2x) .
\end{align}
Since we focus only on the phases with homogeneous singlet
boson and double-hole boson condensates, from now on we set
$s_i=s^\dagger_i=\sigma_i=\sigma^\dagger_i=\bar{s}$ and
$d_i=d^\dagger_i=\delta_i=\delta^\dagger_i=\bar{d}$. Furthermore,
we neglect the possibility of triplet condensation in this paper.

By applying the bond operator representation to all the remaining
terms in the Hamiltonian, the saddle-point Hamiltonian can be
written as follows:

\begin{equation}
H=N\epsilon_{0} +H_\textrm{triplet}+H_\textrm{fermion} ,
\label{H_saddle}
\end{equation}
where the triplet boson Hamiltonian is given by
\begin{align}
H_\textrm{triplet}&=\sum_{\textbf{k}} A_{\textbf{k}} [
t^{\dag}_{\alpha}(\textbf{k})\tau_{\alpha}(\textbf{k})
+\tau^{\dag}_{\alpha}(\textbf{k})t_{\alpha}(\textbf{k}) ]
\nonumber\\
&+\sum_{\textbf{k}} B_{\textbf{k}}
[t^{\dag}_{\alpha}(\textbf{k})\tau^{\dag}_{\alpha}(-\textbf{k})
+\tau_{\alpha}(-\textbf{k})t_{\alpha}(\textbf{k}) ]
\nonumber\\
&+\sum_{\textbf{k}} C_{\textbf{k}}
[t^{\dag}_{\alpha}(\textbf{k})t_{\alpha}(\textbf{k})
+\tau^{\dag}_{\alpha}(\textbf{k})\tau_{\alpha}(\textbf{k}) ] ,
\end{align}
and the single-hole fermion Hamiltonian is given by
\begin{align}
H_\textrm{fermion}=H_\textrm{fermion}^{(0)}+H_\textrm{fermion}^\textrm{interaction},
\end{align}
where
\begin{align}
&H_\textrm{fermion}^{(0)}
\nonumber\\
&=\sum_{\textbf{k}} \Big\{
a_{\textbf{k}}[h^{\dag}_{1a}(\textbf{k})h_{2a}(\textbf{k})
+h^{\dag}_{3a}(\textbf{k})h_{4a}(\textbf{k})] +\textrm{H. c.}
\Big\}
\nonumber\\
&+\sum_{\textbf{k}} b_{\textbf{k}} \Big[
h^{\dag}_{1a}(\textbf{k})h_{1a}(\textbf{k})
+h^{\dag}_{2a}(\textbf{k})h_{2a}(\textbf{k})
\nonumber\\
&\quad\quad\quad+h^{\dag}_{3a}(\textbf{k})h_{3a}(\textbf{k})
+h^{\dag}_{4a}(\textbf{k})h_{4a}(\textbf{k}) \Big],
\end{align}
and
\begin{align}
&H_\textrm{fermion}^\textrm{interaction}
\nonumber\\
&=\sum_{\textbf{k}} \Big\{
c_{\textbf{k}}[h^{\dag}_{3a}(\textbf{k})h_{1a}(\textbf{k})
+h^{\dag}_{2a}(\textbf{k})h_{4a}(\textbf{k})] +\textrm{H. c.}
\Big\} \nonumber \\
&+\sum_{\textbf{k}} \Big\{
d_{\textbf{k}}[h^{\dag}_{4a}(\textbf{k})h_{1a}(\textbf{k})
+h^{\dag}_{2a}(\textbf{k})h_{3a}(\textbf{k})] +\textrm{H. c.}
\Big\} \nonumber \\
&+\sum_{\textbf{k}} \Big\{
e_{1\textbf{k}}[h^{\dag}_{3\uparrow}(\textbf{k})h^{\dag}_{1\downarrow}(-\textbf{k})
+h^{\dag}_{2\uparrow}(\textbf{k})h^{\dag}_{4\downarrow}(-\textbf{k})]
+\textrm{H. c.} \Big\} \nonumber\\
&+\sum_{\textbf{k}} \Big\{
e_{2\textbf{k}}[h^{\dag}_{1\uparrow}(\textbf{k})h^{\dag}_{3\downarrow}(-\textbf{k})
+h^{\dag}_{4\uparrow}(\textbf{k})h^{\dag}_{2\downarrow}(-\textbf{k})]
+\textrm{H. c.} \Big\} \nonumber\\
&+\sum_{\textbf{k}} \Big\{
f_{1\textbf{k}}[h^{\dag}_{4\uparrow}(\textbf{k})h^{\dag}_{1\downarrow}(-\textbf{k})
+h^{\dag}_{2\uparrow}(\textbf{k})h^{\dag}_{3\downarrow}(-\textbf{k})]
+\textrm{H. c.} \Big\} \nonumber\\
&+\sum_{\textbf{k}} \Big\{
f_{2\textbf{k}}[h^{\dag}_{1\uparrow}(\textbf{k})h^{\dag}_{4\downarrow}(-\textbf{k})
+h^{\dag}_{3\uparrow}(\textbf{k})h^{\dag}_{2\downarrow}(-\textbf{k})]
+\textrm{H. c.} \Big\}.
\end{align}
In the above expressions, $N$ is the number of unit cells,
\begin{align}
\epsilon_{0}&=-\frac{3}{2}J\bar{s}^{2}+2\xi(1-\bar{s}^{2}+\bar{d}^{2}-2x)
-2\mu(2x-2\bar{d}^{2})
\nonumber\\
&\quad+J' \left( Q_{x}^{2}+Q_{y}^{2}-P_{x}^{2}-P_{y}^{2} \right)
\nonumber\\
&\quad+\frac{3}{2}J
(|\Pi_{x}|^{2}+|\Pi_{y}|^{2}+4|\Delta_{x}|^{2}+4|\Delta_{y}|^{2})
\nonumber\\
&\quad+V(2+2\bar{d}^{2}-4x)+8V(1-x)^{2}, \nonumber
\end{align}
\begin{align}
A_{\textbf{k}}&=J'( P_{x}\cos{k_{x}} +P_{y}\cos{k_{y}} ),
\nonumber\\
B_{\textbf{k}}&=-J'( Q_{x}\cos{k_{x}} +Q_{y}\cos{k_{y}} ),
\nonumber\\
C_{\textbf{k}}&=J/4 -\xi,
\end{align}
and
\begin{align}
a_{\textbf{k}}&=-t, \quad\quad\quad\quad b_{\textbf{k}}=\mu,
\nonumber\\
c_{\textbf{k}}&=c_{R\textbf{k}}+i c_{I\textbf{k}}, \quad
d_{\textbf{k}}=d_{R\textbf{k}}+i d_{I\textbf{k}},
\nonumber\\
e_{1\textbf{k}}&=e_{R\textbf{k}}+i e_{I\textbf{k}}, \quad
e_{2\textbf{k}}=e_{R\textbf{k}}-i e_{I\textbf{k}},
\nonumber\\
f_{1\textbf{k}}&=f_{R\textbf{k}}+i f_{I\textbf{k}}, \quad
f_{2\textbf{k}}=f_{R\textbf{k}}-i f_{I\textbf{k}},
\label{exactdispersion}
\end{align}
where
\begin{align}
c_{R\textbf{k}}&=-t'\left(\bar{d}^2-\frac{1}{2}\bar{s}^2\right)(\cos{k_{x}}+\cos{k_{y}})
\nonumber\\
&+\frac{t'}{2}(P_{x}\cos{k_{x}}+P_{y}\cos{k_{y}})
\nonumber\\
&-\frac{3}{8}J'(\Pi_{x}\cos{k_{x}}+\Pi_{y}\cos{k_{y}}),
\nonumber\\
c_{I\textbf{k}}&=-t'\left(\bar{d}^2+\frac{1}{2}\bar{s}^2\right)
(\sin{k_{x}}-\sin{k_{x}})
\nonumber\\
&+\frac{t'}{2}(-P_{x}\sin{k_{x}}+P_{y}\sin{k_{y}})
\nonumber\\
&-\frac{3}{8}J'(\Pi_{x}\sin{k_{x}}-\Pi_{y}\sin{k_{y}}), \label{ck}
\end{align}
\begin{align}
d_{R\textbf{k}}&=-t'\left(\bar{d}^2-\frac{1}{2}\bar{s}^2\right)
(\cos{k_{x}}+\cos{k_{y}})
\nonumber\\
&-\frac{t'}{2}(P_{x}\cos{k_{x}}+P_{y}\cos{k_{y}})
\nonumber\\
&-\frac{3}{8}J'(\Pi_{x}\cos{k_{x}}+\Pi_{y}\cos{k_{y}}),
\nonumber\\
d_{I\textbf{k}}&=-t'\left(\bar{d}^2+\frac{1}{2}\bar{s}^2\right)
(\sin{k_{x}}+\sin{k_{y}})
\nonumber\\
&+\frac{t'}{2}(P_{x}\sin{k_{x}}+P_{y}\sin{k_{y}})
\nonumber\\
&-\frac{3}{8}J'(\Pi_{x}\sin{k_{x}}+\Pi_{y}\sin{k_{y}}), \label{dk}
\end{align}
\begin{align}
e_{R\textbf{k}}&=-\sqrt{2}t'\bar{d}\bar{s}(\cos{k_{x}}+\cos{k_{y}})
\nonumber\\
&-\frac{3}{4}J'(\Delta_{x}\cos{k_{x}}+\Delta_{y}\cos{k_{y}}),
\nonumber\\
e_{I\textbf{k}}&=-\frac{3}{4}J'(\Delta_{x}\sin{k_{x}}-\Delta_{y}\sin{k_{y}}),
\label{ek}
\end{align}
and
\begin{align}
f_{R\textbf{k}}&=-\sqrt{2}t'\bar{d}\bar{s}(\cos{k_{x}}+\cos{k_{y}})
\nonumber\\
&-\frac{3}{4}J'(\Delta_{x}\cos{k_{x}}+\Delta_{y}\cos{k_{y}}),
\nonumber\\
f_{I\textbf{k}}&=-\frac{3}{4}J'(\Delta_{x}\sin{k_{x}}+\Delta_{y}\sin{k_{y}}).
\label{fk}
\end{align}
Of course, the order parameters, $P$, $Q$, $\Pi$, and $\Delta$,
should be determined self-consistently by satisfying the following
conditions:
\begin{align}
P_{x} &\equiv \langle t^{\dag}_{{\bf i}\alpha} \tau_{{\bf
i}+\hat{x},\alpha} \rangle, \;\;\; Q_{x} \equiv \langle t_{{\bf
i}\alpha} \tau_{{\bf i}+\hat{x},\alpha} \rangle,
\nonumber\\
\Pi_{x} &= \langle h^{\dag}_{1{\bf i}a} h_{3,{\bf i}-\hat{x},a}
\rangle = \langle h^{\dag}_{1{\bf i}a} h_{4,{\bf i}-\hat{x},a}
\rangle
\nonumber\\
&= \langle h^{\dag}_{2{\bf i}a} h_{3,{\bf i}+\hat{x},a} \rangle =
\langle h^{\dag}_{2{\bf i}a} h_{4,{\bf i}+\hat{x},a} \rangle,
\nonumber\\
\Delta_{x} &= \langle h_{1{\bf i}\downarrow} h_{3,{\bf
i}-\hat{x},\uparrow} \rangle = \langle h_{1{\bf i}\downarrow}
h_{4,{\bf i}-\hat{x},\uparrow} \rangle
\nonumber\\
&= \langle h_{2{\bf i}\downarrow} h_{3,{\bf i}+\hat{x},\uparrow}
\rangle = \langle h_{2{\bf i}\downarrow} h_{4,{\bf
i}+\hat{x},\uparrow} \rangle,
\end{align}
and similar equations for the $y$-direction.

As seen in Eq.~(\ref{exactdispersion}) and subsequent equations,
$c_{\textbf{k}}$, $d_{\textbf{k}}$, $e_{\textbf{k}}$, and
$f_{\textbf{k}}$ are in general complex. In this case it is not
easy to obtain analytic expressions for the $h$-fermion spectrum.
Fortunately, however, it turns out that we can make the following
approximations:
\begin{align}\label{eq:approx}
&c_{\textbf{k}}\cong c_{R\textbf{k}}, \; d_{\textbf{k}}\cong
d_{R\textbf{k}},
\nonumber\\
&e_{1\textbf{k}} \cong e_{R\textbf{k}}, e_{2\textbf{k}} \cong e_{R\textbf{k}},
\nonumber\\
&f_{1\textbf{k}}\cong f_{R\textbf{k}}, f_{2\textbf{k}}\cong
f_{R\textbf{k}}.
\end{align}
The justification for the above approximation will be given in
Appendix \ref{appendix:approx} where we discuss the physical
meaning of the approximation and provide supporting numerical
results. For the time being, however, we proceed to derive
saddle-point equations by accepting Eq.~(\ref{eq:approx}).


\subsection{Self-consistent saddle-point equations}

After the Bogoliubov transformation, the saddle-point Hamiltonian
in Eq.~(\ref{H_saddle}) can be written in the following way:
\begin{align}
H&=N\epsilon_{0}
+\sum_{\textbf{k}}\sum_{l=1}^{4}[b_{\textbf{k}}-\Omega_{l}(\textbf{k})]
\nonumber \\
&+\sum_{\textbf{k}}\sum_{l=1}^{4}\Omega_{l}(\textbf{k})
\gamma^{\dag}_{la}(\textbf{k})\gamma_{la}(\textbf{k}) \nonumber \\
&+\sum_{\textbf{k}}\left\{
w_{\eta1}(\textbf{k})\eta^{\dag}_{1\alpha}(\textbf{k})\eta_{1\alpha}(\textbf{k})
+\frac{3}{2}[w_{\eta1}(\textbf{k})-C_{\textbf{k}}] \right\}
\nonumber\\
&+\sum_{\textbf{k}}\left\{w_{\eta2}(\textbf{k})\eta^{\dag}_{2\alpha}(\textbf{k})\eta_{2\alpha}(\textbf{k})
+\frac{3}{2}[w_{\eta2}(\textbf{k})-C_{\textbf{k}}] \right\},
\end{align}
where
\begin{align}\label{eq:eigenvalues}
w_{\eta1}(\textbf{k})&=\sqrt{(C_\textbf{k}+A_\textbf{k})^{2}-B_\textbf{k}^{2}},
\nonumber\\
w_{\eta2}(\textbf{k})&=\sqrt{(C_\textbf{k}-A_\textbf{k})^{2}-B_\textbf{k}^{2}},
\nonumber\\
\Omega_{1}(\textbf{k})&=\sqrt{(a_\textbf{k}-b_\textbf{k}+c_\textbf{k}-d_\textbf{k})^{2}},
\nonumber\\
\Omega_{2}(\textbf{k})&=\sqrt{(a_\textbf{k}-b_\textbf{k}-c_\textbf{k}+d_\textbf{k})^{2}},
\nonumber\\
\Omega_{3}(\textbf{k})&=\sqrt{(a_\textbf{k}+b_\textbf{k}-c_\textbf{k}-d_\textbf{k})^{2}
+|e_\textbf{k}+f_\textbf{k}|^{2}},
\nonumber\\
\Omega_{4}(\textbf{k})&=\sqrt{(a_\textbf{k}+b_\textbf{k}+c_\textbf{k}+d_\textbf{k})^{2}
+|e_\textbf{k}+f_\textbf{k}|^{2}},
\end{align}
and we set $c_\textbf{k}= c_{R \textbf{k}}$, $d_\textbf{k}= d_{R
\textbf{k}}$, $e_\textbf{k}= e_{R \textbf{k}}$, and $f_\textbf{k}=
f_{R \textbf{k}}$. In the above equations, $\eta_{1\alpha}(\textbf{k})$ and
$\eta_{2\alpha}(\textbf{k})$ describe two bosonic particles
derived from the linear combination of $t_\alpha({\bf k})$ and
$\tau_\alpha({\bf k})$. Similarly, $\gamma_{l}(\textbf{k})$
denote four fermionic Bogoliubov quasi-particles obtained from
$h_{m}(\textbf{k})$ and $h^{\dag}_{m}(-\textbf{k})$ [$m=1,2,3,4$].

Now the ground state energy per unit cell is given by
\begin{align}
\varepsilon_{gr}&= \frac{\langle H \rangle_\textrm{gr}}{N}
\nonumber\\
&=\epsilon_{0}
+\frac{1}{N}\sum_{\textbf{k}}\sum_{l=1}^{4}(b_{\textbf{k}}-\Omega_{l}(\textbf{k}))
\nonumber \\
&+\frac{1}{N}\sum_{\textbf{k}}\frac{3}{2}\left[w_{\eta1}(\textbf{k})-C_{\textbf{k}}\right]
+\frac{1}{N}\sum_{\textbf{k}}\frac{3}{2}\left[w_{\eta2}(\textbf{k})-C_{\textbf{k}}\right].
\end{align}
The four saddle-point equations for $\bar{s}$, $\bar{d}$, $\xi$,
and $\mu$
can be obtained by minimizing the ground state energy as follows:
\begin{align}
\frac{\partial\varepsilon_{gr}}{\partial\xi}=
\frac{\partial(\varepsilon_{gr}/J)}{\partial\bar{s}^{2}}=
\frac{\partial\varepsilon_{gr}}{\partial\mu}=
\frac{\partial(\varepsilon_{gr}/J)}{\partial\bar{d}^{2}}= 0 \;,
\end{align}
where
\begin{align}
&\frac{\partial\varepsilon_{gr}}{\partial\xi}
=5-2\bar{s}^{2}+2\bar{d}^{2}-4x
\nonumber\\
&-\frac{3}{2}\frac{1}{N} \sum_{\textbf{k}} \left[
\frac{C_{\textbf{k}}+A_{\textbf{k}}}{w_{\eta1}(\textbf{k})}
+\frac{C_{\textbf{k}}-A_{\textbf{k}}}{w_{\eta2}(\textbf{k})}
\right], \nonumber
\end{align}
\begin{align}
&\frac{\partial(\varepsilon_{gr}/J)}{\partial\bar{s}^{2}}=-\frac{3}{2}-2\frac{\xi}{J}
\nonumber\\
&+\frac{t'/J}{N}\sum_{\textbf{k}} (\cos{k_{x}}+\cos{k_{y}})
\frac{a_{\textbf{k}}+b_{\textbf{k}}-c_{\textbf{k}}-d_{\textbf{k}}}{\Omega_{3}(\textbf{k})}
\nonumber\\
&-\frac{t'/J}{N}\sum_{\textbf{k}} (\cos{k_{x}}+\cos{k_{y}})
\frac{a_{\textbf{k}}+b_{\textbf{k}}+c_{\textbf{k}}+d_{\textbf{k}}}{\Omega_{4}(\textbf{k})}
\nonumber\\
&+\sqrt{2}\frac{\bar{d}}{\bar{s}} \frac{t'/J}{N}\sum_{\textbf{k}}
(\cos{k_{x}}+\cos{k_{y}}) \left[
\frac{e_{\textbf{k}}+f_{\textbf{k}}}{\Omega_{3}(\textbf{k})}
+\frac{e_{\textbf{k}}+f_{\textbf{k}}}{\Omega_{4}(\textbf{k})}
\right], \nonumber
\end{align}
\begin{align}
&\frac{\partial\varepsilon_{gr}}{\partial\mu}=-4(x-\bar{d}^{2}-1)
\nonumber\\
&-\frac{1}{N}\sum_{\textbf{k}} \Big[
-\frac{a_{\textbf{k}}-b_{\textbf{k}}-c_{\textbf{k}}+d_{\textbf{k}}}{\Omega_{1}(\textbf{k})}
-\frac{a_{\textbf{k}}-b_{\textbf{k}}+c_{\textbf{k}}-d_{\textbf{k}}}{\Omega_{2}(\textbf{k})}
\nonumber\\
&\qquad\qquad\quad+\frac{a_{\textbf{k}}+b_{\textbf{k}}-c_{\textbf{k}}-d_{\textbf{k}}}
{\Omega_{3}(\textbf{k})}
+\frac{a_{\textbf{k}}+b_{\textbf{k}}+c_{\textbf{k}}+d_{\textbf{k}}}
{\Omega_{4}(\textbf{k})} \Big], \nonumber
\end{align}
\begin{align}
&\frac{\partial(\varepsilon_{gr}/J)}{\partial\bar{d}^{2}}=2\frac{\xi}{J}
+4\frac{\mu}{J}+2\frac{V}{J}
\nonumber\\
&-\frac{2t'/J}{N}\sum_{\textbf{k}} (\cos{k_{x}}+\cos{k_{y}})
\frac{a_{\textbf{k}}+b_{\textbf{k}}-c_{\textbf{k}}-d_{\textbf{k}}}{\Omega_{3}(\textbf{k})}
\nonumber\\
&+\frac{2t'/J}{N}\sum_{\textbf{k}} (\cos{k_{x}}+\cos{k_{y}})
\frac{a_{\textbf{k}}+b_{\textbf{k}}+c_{\textbf{k}}+d_{\textbf{k}}}{\Omega_{4}(\textbf{k})}
\nonumber\\
&+\sqrt{2}\frac{\bar{s}}{\bar{d}} \frac{t'/J}{N}\sum_{\textbf{k}}
(\cos{k_{x}}+\cos{k_{y}}) \Big[
\frac{e_{\textbf{k}}+f_{\textbf{k}}}{\Omega_{3}(\textbf{k})}
+\frac{e_{\textbf{k}}+f_{\textbf{k}}}{\Omega_{4}(\textbf{k})}
\Big].
\end{align}

Self-consistency conditions for eight order parameters can also be
written in terms of new quasi-particle spectra:
\begin{align}
&P_{x}=\frac{3}{4}\frac{1}{N}\sum_{\textbf{k}} \cos{k_{x}} \left[
\frac{C_{\textbf{k}}+A_{\textbf{k}}}{w_{\eta1}(\textbf{k})}
-\frac{C_{\textbf{k}}-A_{\textbf{k}}}{w_{\eta2}(\textbf{k})}
\right],
\nonumber\\
&Q_{x}=-\frac{3}{4}\frac{1}{N}\sum_{\textbf{k}} \cos{k_{x}} \left[
\frac{B_{\textbf{k}}}{w_{\eta1}(\textbf{k})}
+\frac{B_{\textbf{k}}}{w_{\eta2}(\textbf{k})} \right],
\nonumber\\
&\Pi_{x}=\frac{1}{4}\frac{1}{N}\sum_{\textbf{k}} \cos{k_{x}}
\frac{a_{\textbf{k}}+b_{\textbf{k}}-c_{\textbf{k}}-d_{\textbf{k}}}{\Omega_{3}(\textbf{k})}
\nonumber\\
&\quad-\frac{1}{4}\frac{1}{N}\sum_{\textbf{k}} \cos{k_{x}}
\frac{a_{\textbf{k}}+b_{\textbf{k}}+c_{\textbf{k}}+d_{\textbf{k}}}{\Omega_{4}(\textbf{k})},
\nonumber\\
&\Delta_{x}= -\frac{1}{8}\frac{1}{N}\sum_{\textbf{k}}\cos{k_{x}}
\left[
\frac{e_{\textbf{k}}+f_{\textbf{k}}}{\Omega_{3}(\textbf{k})}
+\frac{e_{\textbf{k}}+f_{\textbf{k}}}{\Omega_{4}(\textbf{k})}
\right],
\end{align}
and additional four self-consistency conditions for $P_{y}$,
$Q_{y}$, $\Pi_{y}$, and $\Delta_{y}$ are obtained when
$\cos{k_{x}}$ is replaced by $\cos{k_{y}}$.

\subsection{Validity of the approximated $h$-fermion dispersion}
\label{appendix:approx}

In order to justify the approximations used in Eq.(\ref{eq:approx}),
we start by
expressing the fermionic part of the Hamiltonian in terms of the
bonding and anti-bonding fermionic operators that can be
written as follows:
\begin{align}
h_{i+a}& \equiv\frac{1}{\sqrt{2}}(h_{1ia}+h_{2ia}),
\nonumber\\
h_{i-a}& \equiv\frac{1}{\sqrt{2}}(h_{1ia}-h_{2ia}),
\nonumber\\
v_{i+a}& \equiv\frac{1}{\sqrt{2}}(h_{3ia}+h_{4ia}),
\nonumber\\
v_{i-a}& \equiv\frac{1}{\sqrt{2}}(h_{3ia}-h_{4ia}),
\end{align}
where $h_{+}$ and $h_{-}$ describe the bonding and anti-bonding
state for the horizontal dimers, respectively. $v_{+}$ and $v_{-}$
play exactly the same role for the vertical dimers.

\begin{figure}[t]
\centering
\includegraphics[width=6cm]{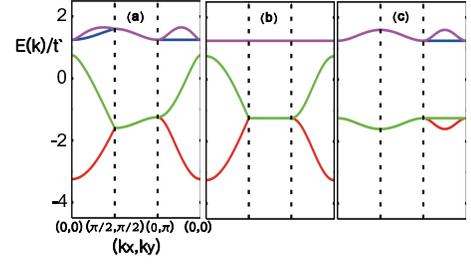}
\caption{(Color online) Band structure of the $h$-fermion at
half-filling. (a) The full band structure of the mean-field
Hamiltonian,
$H_\textrm{fermion}=H_\textrm{fermion}^{(0)}+H_{++,--}+H_{+-,-+}$.
(b) The band structure obtained if $H_\textrm{fermion}$ is
approximated to be $H_\textrm{fermion}^{(0)}+H_{++,--}$. (c) The
additional band structure due to the remaining term, $H_{+-,-+}$.
For convenience, we plot the band structure of
$H_\textrm{fermion}^{(0)}+H_{+-,-+}$. }\label{fig:6}
\end{figure}

The fermionic part of the Hamiltonian can be decomposed into three
parts:
\begin{equation}
H_\textrm{fermion}=H_\textrm{fermion}^{(0)}+H_{++,--}+H_{+-,-+}
\;\;,
\end{equation}
where
\begin{align}
H_\textrm{fermion}^{(0)}&=\sum_{\textbf{k}} a_{\textbf{k}} \Big[
h^{\dag}_{+a}(\textbf{k})h_{+a}(\textbf{k})
-h^{\dag}_{-a}(\textbf{k})h_{-a}(\textbf{k})
\nonumber\\
&\quad\quad\quad\quad+v^{\dag}_{+a}(\textbf{k})v_{+a}(\textbf{k})
-v^{\dag}_{-a}(\textbf{k})v_{-a}(\textbf{k}) \Big]
\nonumber\\
&+\sum_{\textbf{k}} b_{\textbf{k}} \Big[
h^{\dag}_{+a}(\textbf{k})h_{+a}(\textbf{k})
+h^{\dag}_{-a}(\textbf{k})h_{-a}(\textbf{k})
\nonumber\\
&\quad\quad\quad\quad+v^{\dag}_{+a}(\textbf{k})v_{+a}(\textbf{k})
+v^{\dag}_{-a}(\textbf{k})v_{-a}(\textbf{k}) \Big],
\end{align}
\begin{align}
&H_{++,--}
\nonumber\\
&=\sum_{\textbf{k}} c_{R\textbf{k}} \Big[
h^{\dag}_{+a}(\textbf{k})v_{+a}(\textbf{k})
+v^{\dag}_{+a}(\textbf{k})h_{+a}(\textbf{k})
\nonumber\\
&\quad\quad\quad\quad +h^{\dag}_{-a}(\textbf{k})v_{-a}(\textbf{k})
+v^{\dag}_{-a}(\textbf{k})h_{-a}(\textbf{k}) \Big]
\nonumber\\
&+\sum_{\textbf{k}} d_{R\textbf{k}} \Big[
h^{\dag}_{+a}(\textbf{k})v_{+a}(\textbf{k})
+v^{\dag}_{+a}(\textbf{k})h_{+a}(\textbf{k})
\nonumber\\
&\quad\quad\quad\quad -h^{\dag}_{-a}(\textbf{k})v_{-a}(\textbf{k})
-v^{\dag}_{-a}(\textbf{k})h_{-a}(\textbf{k}) \Big]
\nonumber\\
&+\sum_{\textbf{k}} \Big\{
e_{R\textbf{k}}[v^{\dag}_{+\uparrow}(\textbf{k})h^{\dag}_{+\downarrow}(-\textbf{k})
+h^{\dag}_{+\uparrow}(\textbf{k})v^{\dag}_{+\downarrow}(-\textbf{k})
\nonumber\\
&\quad\quad\quad\quad+v^{\dag}_{-\uparrow}(\textbf{k})h^{\dag}_{-\downarrow}(-\textbf{k})
+h^{\dag}_{-\uparrow}(\textbf{k})v^{\dag}_{-\downarrow}(-\textbf{k})]+\textrm{H.
c.} \Big\}
\nonumber\\
&+\sum_{\textbf{k}}
\Big\{f_{R\textbf{k}}[v^{\dag}_{+\uparrow}(\textbf{k})h^{\dag}_{+\downarrow}(-\textbf{k})
+h^{\dag}_{+\uparrow}(\textbf{k})v^{\dag}_{+\downarrow}(-\textbf{k})
\nonumber\\
&\quad\quad\quad\quad-v^{\dag}_{-\uparrow}(\textbf{k})h^{\dag}_{-\downarrow}(-\textbf{k})
-h^{\dag}_{-\uparrow}(\textbf{k})v^{\dag}_{-\downarrow}(-\textbf{k})]
+\textrm{H. c.} \Big\} ,
\end{align}
and
\begin{align}
&H_{+-,-+}
\nonumber\\
&=\sum_{\textbf{k}} i c_{I\textbf{k}} \Big[
v^{\dag}_{-a}(\textbf{k})h_{+a}(\textbf{k})
-h^{\dag}_{+a}(\textbf{k})v_{-a}(\textbf{k})
\nonumber\\
&\quad\quad\quad\quad+v^{\dag}_{+a}(\textbf{k})h_{-a}(\textbf{k})
-h^{\dag}_{-a}(\textbf{k})v_{+a}(\textbf{k}) \Big]
\nonumber\\
&+\sum_{\textbf{k}} i d_{I\textbf{k}} \Big[
v^{\dag}_{+a}(\textbf{k})h_{-a}(\textbf{k})
-h^{\dag}_{-a}(\textbf{k})v_{+a}(\textbf{k})
\nonumber\\
&\quad\quad\quad\quad\quad+h^{\dag}_{+a}(\textbf{k})v_{-a}(\textbf{k})
-v^{\dag}_{-a}(\textbf{k})h_{+a}(\textbf{k}) \Big]
\nonumber\\
&+\sum_{\textbf{k}} \Big\{ i
e_{I\textbf{k}}[v^{\dag}_{-\uparrow}(\textbf{k})h^{\dag}_{+\downarrow}(-\textbf{k})
-h^{\dag}_{+\uparrow}(\textbf{k})v^{\dag}_{-\downarrow}(-\textbf{k})
\nonumber\\
&\quad\quad\quad\quad+v^{\dag}_{+\uparrow}(\textbf{k})h^{\dag}_{-\downarrow}(-\textbf{k})
-h^{\dag}_{-\uparrow}(\textbf{k})v^{\dag}_{+\downarrow}(-\textbf{k})]
+\textrm{H. c.} \Big\}
\nonumber\\
&+\sum_{\textbf{k}} \Big\{ i
f_{I\textbf{k}}[v^{\dag}_{+\uparrow}(\textbf{k})h^{\dag}_{-\downarrow}(-\textbf{k})
-h^{\dag}_{-\uparrow}(\textbf{k})v^{\dag}_{+\downarrow}(-\textbf{k})
\nonumber\\
&\quad\quad\quad\quad+h^{\dag}_{+\uparrow}(\textbf{k})v^{\dag}_{-\downarrow}(-\textbf{k})
-v^{\dag}_{-\uparrow}(\textbf{k})h^{\dag}_{+\downarrow}(-\textbf{k})]
+\textrm{H. c.} \Big\},
\end{align}
where it is important to note that $H_{++,--}$ describes the
fermion hopping and pairing processes within the band of the same
bonding character. On the other hand, $H_{+-,-+}$ contains mixing
processes between the bonding and anti-bonding bands. The
approximation used in Eq.(\ref{eq:approx}) corresponds to the
omission of $H_{+-,-+}$ from $H_\textrm{fermion}$.

To judge the validity of this approximation, let us first consider
the effect of hopping terms alone from the Hamiltonian. In
Fig.~\ref{fig:6}~(a) we plot the full band structure of
$H_\textrm{fermion}$ at half filling. Since the unit cell consists
of four spins, there are four bands. The two bands lying high in
energy are almost flat while the other two are dispersive. In
Fig.~\ref{fig:6}~(b) we show the band structure obtained when
$H_\textrm{fermion}$ is approximted to be
$H_\textrm{fermion}^{(0)}+H_{++,--}$. As one can see, the
difference between the full and the approximated band structure is
minor. The almost flat two bands in the full Hamiltonian, which
are high in energy and so do not play important roles at small
hole doping, become completely localized. The shapes of the other
two dispersive bands are only slightly modified. Now, to
explicitly demonstrate the minor influence of the remaining part,
i.e., $H_{+-,-+}$, we consider an extreme situation where the
role of $H_{+-,-+}$ is maximized. That is, we consider the band
structure of $H_\textrm{fermion}^{(0)}+H_{+-,-+}$. As seen in
Fig.~\ref{fig:6}~(c), the resulting band structure is almost
featureless. Thus, the main effect of $H_{+-,-+}$ would be
providing only a weak distortion to the bands of $H_{++,--}$.

Since normal mixing between the bands with different bonding
characters generates minor effects, it is reasonable to assume
that the pairing process between different bonding bands also does
not modify the band structure significantly. It can be further
argued that these results are valid also for the case of finite
doping because the fermionic band structure has
similar structure to the half-fillied case as can be seen in
Eq.~(\ref{eq:eigenvalues}) and there is no level crossing upon
doping. In conclusion, we believe that the approximation used in
Eq.(\ref{eq:approx}) is accurate enough to capture the essential
physics while it renders convenient simplification in
calculations.



\end{document}